\documentclass[12pt]{article}
\usepackage{graphicx}
\usepackage{epsfig}
\usepackage{amsbsy}

\begin{document}
\baselineskip 10mm\centerline{\large \bf On the Thermal Stability of Graphone}

\vskip 6mm

\centerline{A. I. Podlivaev  and L. A. Openov$^{*}$}

\vskip 4mm

\centerline{\it National Research Nuclear University {}``MEPhI{}'', 115409 Moscow, Russia}

\vskip 2mm

$^{*}$ E-mail: LAOpenov@mephi.ru

\vskip 8mm

\centerline{\bf ABSTRACT}

Molecular dynamics simulation is used to study thermally activated migration of hydrogen atoms
in graphone, a magnetic semiconductor formed of a graphene monolayer with one side covered with
hydrogen so that hydrogen atoms are adsorbed on each other carbon atom only.
The temperature dependence of the characteristic time of disordering of
graphone via hopping of hydrogen atoms to neighboring carbon atoms is established directly. The
activation energy of this process is found to be $E_a=(0.05\pm0.01)$ eV. The small value of $E_a$
points to extremely low thermal stability of graphone, this being a serious handicap for practical
use of the material in nanoelectronics.

\vskip 5mm

\newpage

\centerline{\bf 1. INTRODUCTION}

Graphene [1], a hexagonal monolayer of carbon atoms, is of great interest for both fundamental
physics (Dirac fermions in solids) and practical applications (nanoelectronics [2]). At present,
attention is being attracted to various derivatives of graphene. Among these are such as, e.g.,
graphane CH, a graphene monolayer with both sides completely saturated with hydrogen [3, 4]; diamane,
double-layered graphene with hydrogen atoms adsorbed on its external surfaces [5]; etc.

Recently [6], the existence of another graphene-based hydrocarbon, graphone C$_2$H, was predicted.
Graphone is a graphene monolayer on which hydrogen atoms are adsorbed at only one side (rather than
at both sides, as in graphane), being attached to carbon atoms of one graphene sublattice.
According to the DFT calculations performed in Ref. [6], graphone, like
graphane, is an insulator, but with the much narrower band gap $E_g\approx 0.5$ eV (in graphane,
$E_g\approx 5$ eV [7]). It is of interest that, as follows from the theory [6], graphone exhibits
magnetic properties, in contrast to nonmagnetic graphene and graphane. The local magnetic moments
of nonhydrogenated carbon atoms are ordered ferromagnetically at the Curie temperature $T_C=$
300-400 K. Due to this property, graphone may offer considerable promise for use in various
nanoelectronic (e.g., spintronic) devices.

The range of possible applications of graphone depends to a large extent on the degree of its
thermal stability. In fact, thermally activated hopping of hydrogen atoms between neighboring
carbon atoms can induce substantial structural distortions of graphone and, as a consequence,
uncontrollable changes in its magnetic characteristics (including their irregular distribution
over the sample and even total loss of magnetic properties). In Ref. [6], the stability of graphone
at the temperature $T=300$ K was demonstrated by DFT molecular dynamics simulation. However, the
time of the {}``computer experiment'' during which graphone retained its structure was only 3 ps.
It is clear that this time is not long enough for final inferences on the stability of graphone
to be made.

It is known that the ab initio calculations in simulations of dynamic processes demand very heavy
computational resources and, therefore, allow one to study the evolution of a system of $\sim$100 atoms
(a cluster or a supercell with periodic boundary conditions) over a rather short time ($\sim$10 ps)
inadequate to provide the sufficient statistics. The goal of this study is to implement molecular
dynamics simulation of thermally activated migration of hydrogen in graphone in the tight-binding
model [8] that presents a rational compromise between the stricter ab initio approaches and
the oversimplified approaches involving classic potentials of interatomic interaction. The
tight-binding model adequately describes both small-sized carbon clusters [9-13] and macroscopic
systems [8]. In combination with the molecular dynamics method, the model allows one to bring the
time of simulation up to 1 ns - 1 $\mu$s [9-14]. Previously, this model was successfully used to study
thermal desorption of hydrogen from graphane [13] and the effect of spontaneous regeneration of the
disordered graphene/graphane interface [14] as well as to calculate the dependence of the band
gap of graphane nanoribbons on their width [15]. We show that the activation energy of hopping of
hydrogen atoms between neighboring carbon atoms in graphone is extremely low (almost two orders
of magnitude lower than the activation energy of hydrogen desorption) and, as a result, the
characteristic time of disordering of the graphone structure is no longer than 1 ns even at liquid
nitrogen temperature.

\vskip 10mm

\centerline{\bf 2. METHODS OF CALCULATIONS}

The molecular dynamics simulation of thermally activated migration of hydrogen in graphone was
implemented for the C$_{54}$H$_{60}$ cluster. This cluster represents a graphene fragment, the
edges of which are passivated with hydrogen (the number of passivating peripheral hydrogen atoms
is 48); in addition, the fragment contains 12 {}``inner'' hydrogen atoms adsorbed at one of its
sides, thus forming a graphone fragment (Fig. 1). Passivation is required to saturate dangling
bonds of the  sp- and  sp$^2$-hybridized edge carbon atoms and, thus, first, to reduce the
effects of finite dimensions and, second, to exclude (or minimize) hopping of hydrogen atoms
from inner carbon atoms to edge ones.

At the initial point in time, random velocities and displacements were imparted to each atom (so
that the momentum and angular momentum of the cluster as a whole are zero). Then the forces acting
on the atoms were calculated and the classical equations of motion were solved numerically, with
the step in time $t_{0}=2.72\times 10^{-16}$ s. In the process of simulation, the total energy of
the system remained unchanged (a microcanonical ensemble [16, 17]), so that the role of temperature
was played by the so-called dynamic temperature, the measure of energy of relative motion of atoms.
The dynamic temperature was calculated by the formula [17, 18]
$\langle E_{\textrm{kin}} \rangle=\frac{1}{2}k_{B}T(3n-6)$, where
$\langle E_{\textrm{kin}} \rangle$ is the time-averaged kinetic energy of the system, $k_B$ is
the Boltzmann constant, and $n=$114 is the number of atoms in the cluster (corrections for the
finite dimensions of the thermal reservoir [19, 20] were disregarded because of the large number
of atoms, $n>$100).

To calculate the forces acting on atoms at each step of molecular dynamic simulation,
we used the nonorthogonal tight-binding model [8] modified from the model used in Ref. [21].
This model explicitly includes the quantum-mechanical ({}``band'') contribution of the electron
subsystem to the total energy. We took into account all valence electrons of the system, one
electron of each hydrogen atom (1$S$) and four electrons of each carbon atom (2$S$ and 2$P$).

To calculate the height $U$ of the energy barrier that hampers migration of hydrogen over graphone,
we investigated the hypersurface of the potential energy of the system $E_{pot}$ as a function of
coordinates of the constituent atoms. The stationary points of $E_{pot}$ (local minimums and saddle
configurations) were determined by the method of structural relaxation and by searching in normal
coordinates [22, 23].

\vskip 10mm

\centerline{\bf 3. RESULTS}

By analyzing atomic configurations created in the course of simulation of graphone dynamics, we have
found directly the time of hydrogen migration $\tau$ for 66 different sets of initial atomic
velocities and displacements corresponding to the temperatures $T=$50-400 K. The time $\tau$ was
determined as the time interval from the beginning of simulation to the hop of 1 of 12
nonpassivating hydrogen atoms (Fig. 1) to a neighboring carbon atom. As a result of such a hop,
one covalent C-H bond breaks and another bond is formed. The process of bond {}``switching''
occurs within a time of $\sim$10 fs. No reverse hop of the migrating atom has been ever observed;
i.e., in graphone, migration of hydrogen is an irreversible process, as distinct from migration
in the graphane/graphene structure [14].
The physical reason of such irreversibility is that migration lowers energy;
i.e., graphone presents a metastable configuration (corresponding to a local energy minimum rather
than to the global minimum) and the hydrogen atoms gain energy if they form bonds with neighboring
carbon atoms belonging to different graphene sublattices. We will return to this issue later.

Figure 2 shows the calculated dependence of the logarithm of $\tau$ on inverse temperature.
From Fig. 2, it is evident that this dependence can be rather adequately approximated with a
straight line, suggesting that the standard Arrhenius formula is applicable to the description
of hydrogen migration:
\begin{equation}
\tau^{-1}(T)=A\exp\left(-\frac{E_a}{k_{B}T}\right)~.
\label{1}
\end{equation}
Here, $A$ is the frequency factor independent of (or slightly dependent on) temperature and
$E_a$ is the migration activation energy determined from the slope of the straight line in Fig. 2.
As the temperature is lowered from 400 to 50 K, the time of migration $\tau$ exponentially
increases by four orders of magnitude, from $\sim$0.1 ps to $\sim$1 ns. Statistical analysis of the results of
the computer experiment yields $E_a=(0.05\pm 0.01)$ eV and $A=10^{13.5\pm 0.1}$ s$^{-1}$.
We draw attention to the very small value of $E_a$ that is about 50 times lower than the
activation energy of hydrogen desorption from graphane [13]. Physically, this difference arises
from the fact that desorption necessitates the breakdown of the strong covalent C-H bond, whereas
migration occurs if the C-H bond only {}``switches over'' from one carbon atom to another.

Since the activation energy of migration is defined by the height $U$ of the barrier that hampers
the migration process, we have calculated the value of $U$ in graphone. This was done for a graphone
C$_{200}$H$_{100}$ supercell composed of 100 C$_2$H unit cells with periodic boundary conditions.
The dependence of the potential energy of the system on the reaction coordinate is shown in Fig. 3.
From Fig. 3, it can be seen that the barrier for migration of a hydrogen atom is very low,
$U=0.058$ eV, in agreement with the low activation energy (as a rule, $U\approx E_a$ [10]). In the
coordinate space, this barrier is closer to the carbon atom forming the C-H bond that breaks upon
migration than to the carbon atom involved in the new C-H bond. It should be emphasized that
migration results in a sharp (by 1.46 eV) decrease in energy and, therefore, the barrier for
reverse hopping of the hydrogen atom is very high, see Fig. 3. This is the cause of the above
mentioned irreversibility of the migration process.

\vskip 10mm

\centerline{\bf 4. DISCUSSION}

According to the calculations, the time of migration of hydrogen atoms in graphone at $T=300$ K is
$\tau=$ 0.1-1 ps (the rather large spread of the values of $\tau$ at a particular temperature is
due to the fact that the process of migration is probabilistic in nature and, therefore, the value
of $\tau$ can be markedly different for different sets of initial atomic displacements and
velocities, even if these sets correspond to the same temperature). At first glance, this seems in
contradiction with the results of Ref. [6], wherein the authors also used molecular dynamics
simulation and showed that, at $T=300$ K, graphone retained its structure for 3 ps. However, it
should be noted that the authors of Ref. [6] reported the results of only one
{}``computer experiment'' at one temperature,
these data being insufficient for an unambiguous conclusion to be drawn on the degree of thermal
stability (particularly because of the above-mentioned probabilistic character of the process of
migration). In contrast, in this study, we have carried out a comprehensive analysis of the dynamics
of graphone in a wide temperature range and, on this basis, we have gained a rather large set of
statistics and determined the temperature dependence of the migration time. One more factor that
could influence the results obtained in Ref. [6] was the overly small dimensions of the system
(C$_8$H$_4$) for which the simulation was implemented.

It is worth noting that the results obtained in this study for the activation energy of
hydrogen migration and for the height of the barrier hampering the migration, $E_a=0.05\pm 0.01$ eV
and $U=0.06$ eV, are in excellent agreement with the value of $U=0.06$ eV determined for graphone
in Ref. [24] by ab initio calculations. In addition, the lowering of the energy of graphone on
migration of one hydrogen atom over a distance corresponding to the C-C bond length (1.46 eV) is
very close to the value of 1.44 eV obtained in Ref. [24].

We believe that the extremely low degree of thermal stability of graphone makes this nanocarbon
material unpromising for practical use in electronic devices. In fact, as follows from the estimates
obtained by the above formula and from the above-determined parameters $A$ and $E_a$ entering in
this formula, the characteristic time of structural disordering of graphone becomes macroscopic
($\sim$10$^{10}$ s), only if graphone is cooled down to $T\approx 10$ K. At higher temperatures (Fig. 1),
because of migration of hydrogen, the atomic configuration of graphone breaks down very quickly
(the time of migration is no longer than 1 ns even at $T=$77 K). This can lead to the irregularity
in the distribution of electronic and magnetic characteristics over the sample. In addition, the
possibility of phase separation of graphone into regions enriched with and depleted of hydrogen
should not be ruled out. This issue calls for further investigations.

\vskip 10mm

\centerline{\bf 5. CONCLUSIONS}

It is inferred that, in contrast to graphene and graphane, graphone is an inappropriate material
for use in nanoelectronics because of its low stability even at liquid-nitrogen temperature. An
alternative to graphone as a nanocarbon magnetic semiconductor may be represented by, e. g.,
fluorinated graphene [24], wherein fluorine is substituted for hydrogen.

\vskip 10mm

\centerline{\bf ACKNOWLEDGMENTS}

We thank M.M. Maslov for his help in carrying out this study and discussions of the results.

\newpage
\vskip 2mm
\includegraphics[width=\hsize,height=15cm]{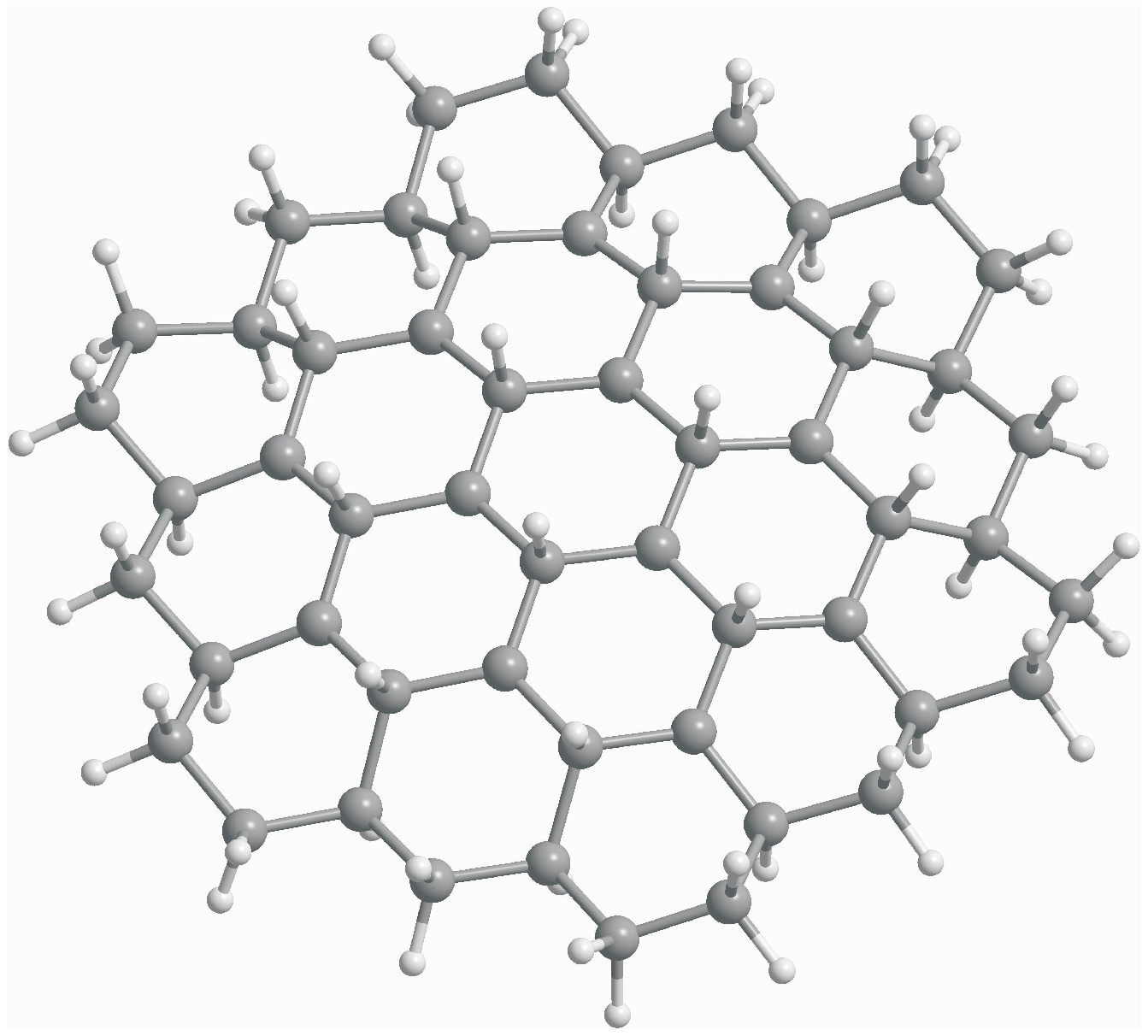}
\vskip 20mm
Fig. 1. The C$_{54}$H$_{60}$ cluster as a fragment of graphone. The large and small
balls are carbon and hydrogen atoms, respectively.

\newpage
\vskip 2mm
\includegraphics[width=\hsize,height=15cm]{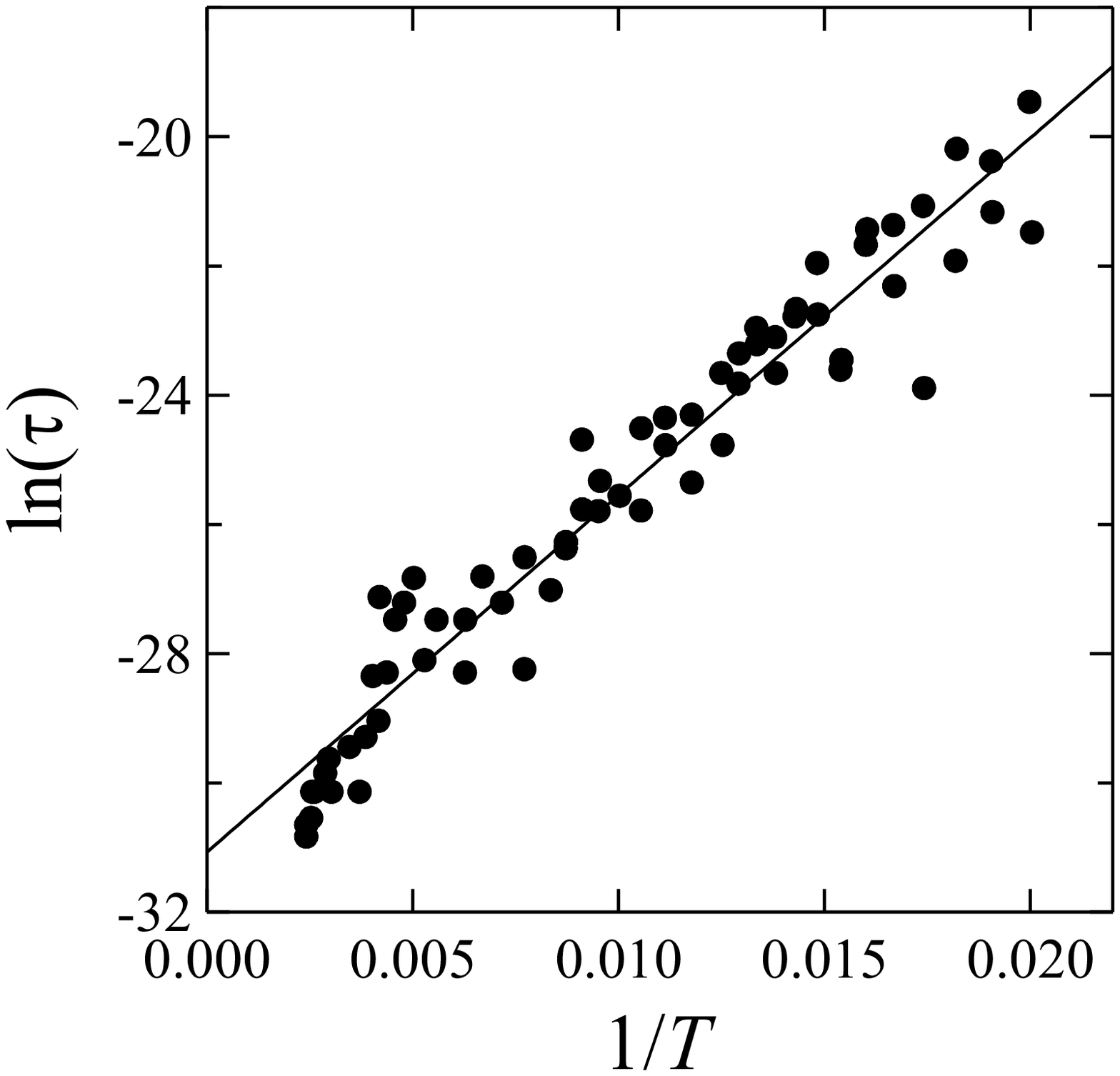}
\vskip 20mm
Fig. 2. The dependence of the logarithm of the migration time $\tau$ (sec) for one hydrogen atom in
the C$_{54}$H$_{60}$ cluster on the inverse temperature $T$ (K$^{-1}$). Symbols refer to the results
of calculations; the solid line is the linear approximation by the least squares method.

\newpage
\vskip 2mm
\includegraphics[width=\hsize,height=15cm]{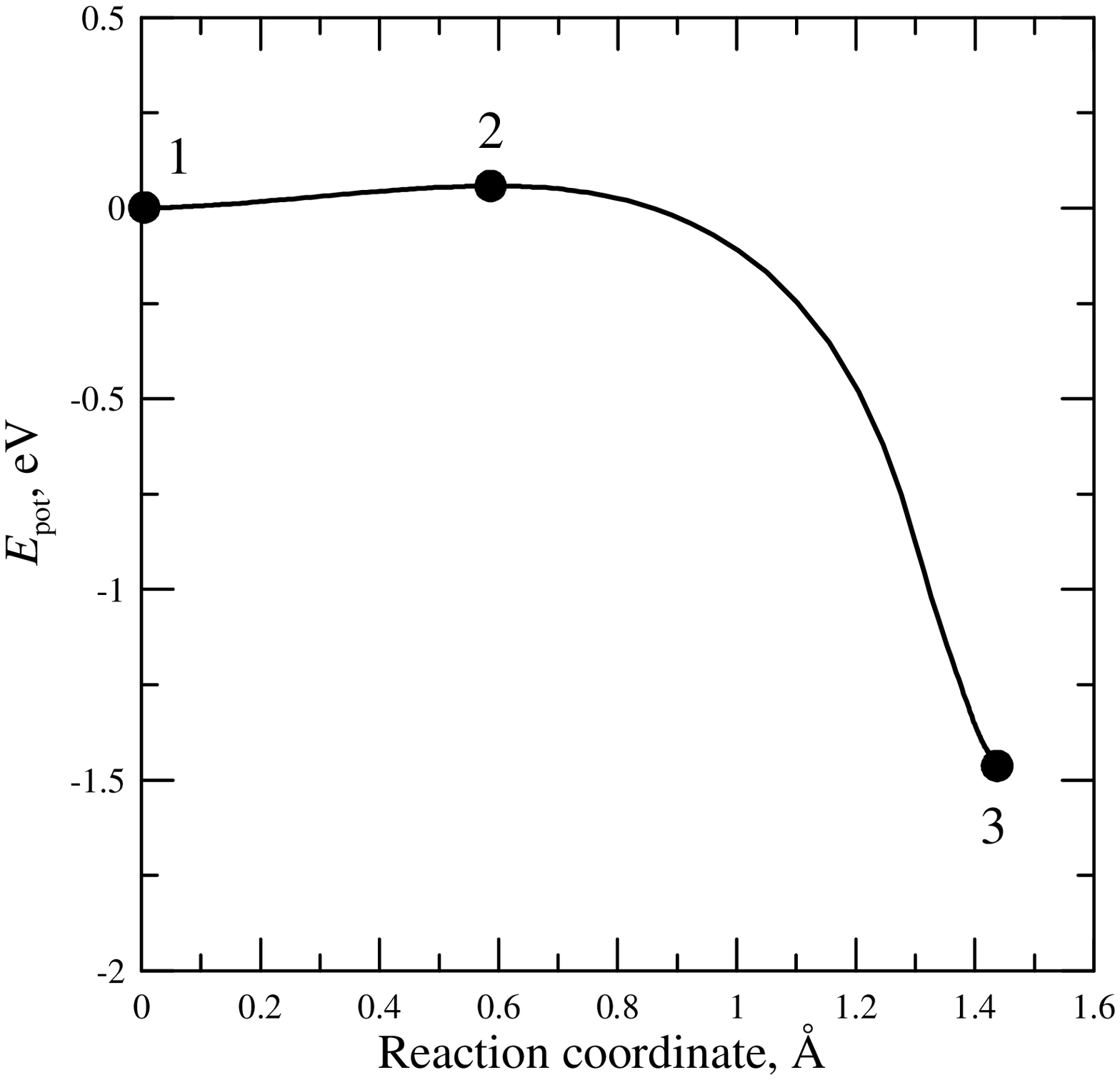}
\vskip 20mm
Fig. 3. The dependence of the potential energy $E_{pot}$ of the C$_{200}$H$_{100}$ supercell with
periodic boundary conditions on the reaction coordinate for migration of a hydrogen atom over a
distance equal to the C-C bond length. For the origin, the energy of the initial atomic
configuration (before migration) is taken. The reaction coordinate is a straight line passing
through two carbon atoms, between which the hydrogen atom migrates. Circles refer to (1) the
initial configuration of graphone, (2) the saddle point that defines the energy barrier for the
migration process, and (3) the configuration after the migration.

\end{document}